\documentclass[12pt,letterpaper]{article}
\usepackage[body={18cm, 25cm}]{geometry}
\usepackage{paralist}
\usepackage{multicol}
\usepackage{multirow}
\usepackage{float}
\usepackage{amsmath}
\usepackage{amsfonts}
\usepackage{amssymb}
\usepackage{graphicx}
\usepackage{geometry}
\usepackage [round]{natbib}
\usepackage{subfig}
\usepackage[pdftex, pdftitle={Diffraction}, pdfauthor={Nicolas Guarin, Juan Gomez}, pdfsubject={Site effects in earthquake engineering}, pdfkeywords={Scattering,Site effects,Transfer function, Wave propagation}, pdfpagemode=UseOutlines,bookmarks,bookmarksopen,pdfstartview=FitH,colorlinks,linkcolor=blue, urlcolor=black, citecolor=blue]{hyperref} 
\usepackage{cleveref}

\begin{document}
\setlength{\parskip}{6mm}
\title{\textbf{Seismic Wave Scattering Through a Compressed Hybrid BEM/FEM Method.} }
\author{Nicol\'as Guar\'in-Zapata, Juan Gomez and Juan Jaramillo\\
        Departamento de Ingenier\'ia Civil\\
                Universidad EAFIT\\
                Medell\'in,\\
                Colombia\\
                \texttt{jgomezc1@eafit.edu.co}}  
\date{\today}
\maketitle
\begin{abstract}
Approximate numerical techniques, for the solution of the elastic wave scattering problem over semi-infinite domains are reviewed.  The approximations involve the representation of the half-space by a boundary condition described in terms of 2D boundary element discretizations. The classical BEM matrices are initially re-written into the form of a dense dynamic stiffness matrix and later approximated to a banded matrix.  The resulting final banded matrix is then used like a standard finite element to solve the wave scattering problem at lower memory requirements. The accuracy of the reviewed methods is benchmarked against the classical problems of a semi-circular and a rectangular canyon.  Results are presented in the time and frequency domain, as well as in terms of relative errors in the considered approximations.  The main goal of the paper is to give the analyst a method that can be used at the practising level where an approximate solution is enough in order to support engineering decisions.

\end{abstract}
\section{Introduction}
\label{sec:intro}
One of the main challenges in the numerical simulation of wave scattering problems, over infinite or semi-infinite domains, is the proper imposition of radiation boundary conditions: When the analysis is performed using a full domain method, e.g., the finite differences method (FDM) or the finite element method (FEM), one has to render finite the computational domain, and use artificial boundary elements to approximate radiation damping.  A wide variety of absorbing boundary elements are available in the literature, \citep{lysmer1969finite, engquist1977absorbing, clayton1977absorbing}. Most of these elements suffer deficiencies, like dependence of the performance on the angle of the incident wave or a limited frequency band of efficient operation. On the other hand, among the many techniques available to approximate Sommerfeld's radiation boundary conditions, those based on integral formulations are the most accurate \citep{vara1976, banerjeeboundary, sanchez2006indirect}; although their coupling to existing finite element codes introduces inconvenient features, since their resulting coefficient matrices are asymmetric and fully populated and destroy many of the appealing features of finite element methods. In this work we explore approximations to the radiation boundary condition, that combine the accuracy of integral-based formulations with the advantages of the finite element algorithm. Our approximate representation is intended to be useful in the solution of scattering problems of moderate size using existing finite element codes and the resources available in current personal computers.

In general, the complexity of a formulation to approximate the radiation boundary condition is a function of the degree of space-time non-locality introduced in the boundary model. In the case of classical FDs and FEs, non-locality is an unnatural condition: Consequently, in order to obtain effective solutions, the analyst must sacrifice accuracy at the expense of non-locality; with accurate results obtained after a trade-off between non-locality and the use of Saint-Venant-end-effects leading to large computational models. One of the first contributions in absorbing boundaries, but still one of the most popular ones, is the fully-local, viscous boundary element by \cite{lysmer1969finite}. This boundary provides the conceptual basis of most of the currently used elements, which exhibit various levels of efficiency with increasing levels of non-locality, \citep{berenger1994perfectly, soudkhah2012wave, lee2012dynamic, zheng2013anisotropic, duru2012well}. Moreover, many of these absorbing conditions have been used in recent FDs and FEs simulations of seismic regions, as large as the state of California and for frequencies up to $4.0$ Hz, \citep{frankel1993three, komatitsch1999, min2003, bielak2003, komatitsch2004, givoli2004high, frehner2008, ichimura2009, lee2009b, lee2009a, chul-ho2010,  bielak2010, chaljub2010, lan2011}. In a second class of methods, which are inherently non-local in space and time, \citep{vara1976, book:BEMElastodynamics, sanchez-sesma91, banerjeeboundary, sanchez-sesma95, janod2000seismic, iturraran-viveros2005}, the radiation condition is imposed through fully coupled boundary integral formulations: In those cases the half-space condition is analytically satisfied by the Green's function, but the resulting algorithm, although highly accurate is also computationally demanding with coefficient matrices which are non-symmetric and fully populated.

A third solution strategy that has been popular during many years, is a combination that approximates radiation damping via the BEM technique, while the scatterer, (including heterogeneities and/or nonlinearities), is treated with an FEM algorithm \citep{brebbia79, beer81, book:BEMBanerjee, zienkiewicz_coupling77, book:zienkiewicz_FEM2, boutchicha2007, bem-abaqus, dam-reservoir2009, bielak2009}. That approach becomes especially attractive when the BEM discretization is written in  the form of a displacements-based finite element, representing the semi-infinite domain as a half-space-super-element (HSSE) and favouring its straightforward coupling to a finite element model for the scatterer. Although the coupling of BEM and FEM algorithms can be accomplished in a wide majority of commercially available finite element codes that assimilate user element subroutines, e.g., \citep{abaqus1989karlsson, FEAP_programmer_manual}, the asymmetric and dense character of the HSSE-coefficient-matrix still restrains the advantages of the finite element method. Therefore, additional modifications must be imposed to the resulting radiation boundary approximation, in order to obtain an efficient and accurate formulation that preserves the nice properties of the finite element method. In this work we enforce the HSSE-coefficient-matrix to remain banded and symmetric, and asses the effect upon the solution accuracy of such compression scheme.

Compression schemes for the boundary element matrices have been used in the past, but all of them performed the matrix compression at the boundary element level,i.e. in the discrete versions of the displacements and traction kernels. For instance, \cite{bouchon-herod95} used the criteria of making zero each term of the traction and displacement matrices, where an original absolute value  below a preselected threshold was identified; \cite{Hmatrices-2006} used hierarchical matrices and adaptive cross approximation, which is a purely algebraic method, to compress the discrete BEM integral operators; \cite{abe2001} and \cite{eppler2005} used a wavelet BEM technique to achieve the packing of the resulting matrices. The fast multi-poles boundary element methods (FM-BEM) introduced by \cite{book:liu_FMBEM}, have gained popularity due to its low computational cost in both, storage requirements and CPU time; these methods are based on a far field expansion of the kernels of the integral equations. In this work, we follow a slightly different approach since we directly operate in the final resulting BEM dynamic stiffness matrix; this operation is a mandatory requirement if the goal is to preserve the advantages of the classical displacement-based finite element method. Such requirement is clarified if one considers the fact that the inverse of a banded matrix is not generally banded, and although the product of banded matrices is still banded, the resulting half-band-width is larger than in the original single matrices.

In the current work we first review the well-known direct and indirect BEM algorithms, and establish the connection between the final HSSE matrices obtained with both methods. The resulting matrices are then compressed using (i) a threshold criteria and (ii) a method based on the distance to the diagonal of the different terms; this method is termed herein a half-band-width-method. The resulting HSSEs are coupled into an existing finite element code and tested in the solution of two well documented scattering problems, namely a semicircular and a rectangular canyon. Both problems are rich in scattered motions and contain different sources of diffraction: As shown by different authors \citep{bielak2006effective, jaramillo2013analytic} the scattered and diffracted parts of the total field contain the motions that must be absorbed by the HSSE. The article is organized as follows: The first part summarizes the scattering problem in terms of integral equations, leading to boundary element algorithms and the formulation of the HSSE coefficient matrices, together with its coupling to existing finite element codes. We then discuss and apply the compression criteria to the solution of both canyons. The results, which are shown in the frequency and time domain are then compared in terms of transfer functions and synthetic seismograms along the canyon surfaces.

\section{Solution of the Scattering Problem}
\subsection{Direct and Indirect BEM Formulation and coupling to the finite element method}
\begin{figure}[H]
\centering
\includegraphics[width=3.0 in]{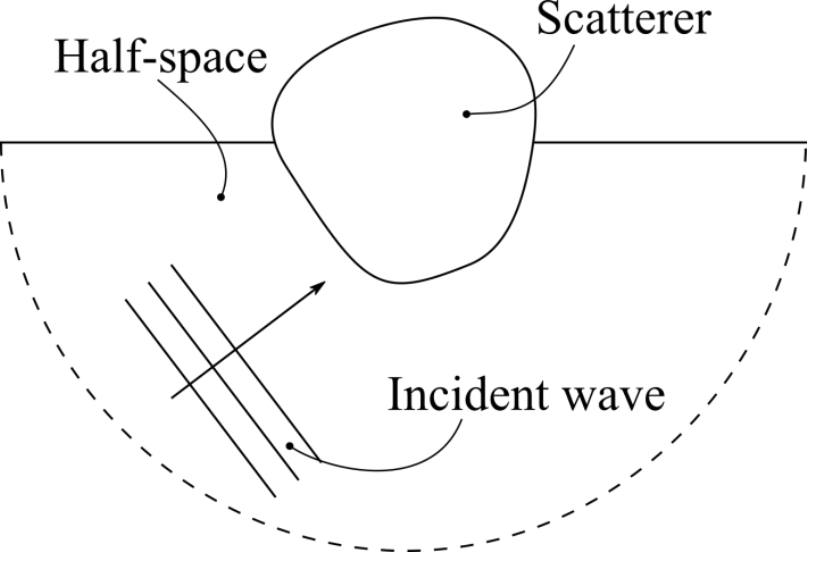}
\caption{Schematic representation of the scattering problem.}
\label{fig:scattering domain}
\end{figure}

The general scattering problem is schematized in \cref{fig:scattering domain}.  Since the scatterer is treated with a classical FEM algorithm, we only discuss the formulation of the boundary value problem for the half-space part of the domain.  In this work we use a method due to \cite{bielak2006effective} where the only unknown in the half-space corresponds to the scattered field.  In this way, the external plane wave excitation is converted into an equivalent system of internal sources, located along the contact interface between the two media.  The problem is then formulated in terms of the Green's function using either an integral representation theorem or a plain superposition of sources.

Regardless of whether a direct or indirect BEM approach is used, both methods find their basic foundation in the concept of the fundamental solution or the specific problem Green's functions $G_{ij}(\vec{x},\vec{\xi})$ and $H_{ij}(\vec{x},\vec{\xi},\hat{n}_x)$. These functions allow us to express respectively the  $i-th$ displacement $u_i$ and traction components $t_i$ at a field point $\vec{x}$ due to a load $P_j$ applied in the $j-th$ direction at a point $\vec{\xi}$ according to \cref{Green};

\begin{equation}\label{Green}
\begin{aligned}
u_i(\vec{x})&=G_{ij}\left(\vec{x},\vec{\xi}\right)P_j\left(\vec{\xi}\right)\\
t_i(\vec{x},\hat{n}_x)&=H_{ij}\left(\vec{x},\vec{\xi},\hat{n}_x\right)P_j\left(\vec{\xi}\right).
\end{aligned}
\end{equation}

and where it is clear that $H_{ij}(\vec{x},\vec{\xi},\hat{n}_x)$ is the tractions counterpart of the displacement Green's function $G_{ij}(\vec{x},\vec{\xi})$. In the above $\hat{n}_x$ denotes the normal vector outward to the surface at a field point $\vec{x}$. Direct use of \cref{Green} into Betti's reciprocity theorem for the scattered and the fundamental states, yields Somigliana identity which is the basis of the integral equation to be solved in a direct boundary element method as given in \cref{somigliana}

\begin{equation} \label{somigliana}
C_{ij}(\vec{\xi}) u_i(\vec{\xi})=\int_S G_{ij}(\vec{x},\vec{\xi}) t_i(\vec{x},\hat{n}_x)\,\mathrm{d}S(\vec{x})-\int_S H_{ij}(\vec{x},\hat{n}_x;\vec{\xi},\hat{n}_\xi) u_i(\vec{x})\,\mathrm{d}S(\vec{x}).
\end{equation}

where again $\hat{n}_x$ is the surface outward normal unit vector; $C_{ij}$ is a coefficient that depends on the smoothness of the boundary; $u_i$ and $t_i$ are the boundary  displacements and tractions vectors respectively. Similarly, direct application of \cref{Green} after assuming that the solution field is produced by an arbitrary distribution of source densities $\phi_j(\vec{\xi})$, along the boundary of the considered domain, yields the fundamental equations to be solved in an indirect boundary element method, \cref{IBEM};

\begin{equation}\label{IBEM}
\begin{aligned}
u_i(\vec{x})&=\int_S G_{ij}(\vec{x},\vec{\xi}) \phi_j(\vec{\xi})\,\mathrm{d}S(\vec{\xi})\\
t_i(\vec{x},\hat{n}_x)&=\dfrac{1}{2}\phi_j(\vec{x})+\int_S H_{ij}(\vec{x},\vec{\xi},\hat{n}_\xi) \phi_j(\vec{\xi})\,\mathrm{d}S(\vec{\xi})
\end{aligned}
\end{equation}

and where the problem is now solved in a two step algorithm, where one has to find the distribution of source densities $\phi_j(\vec{\xi})$ that solves the specified boundary conditions and then finds the solution field anywhere inside the domain.

The corresponding system of algebraic equations in either case can be obtained after discretization of the field and the boundary into $N$ (constant) elements.  In the case of the direct BEM approach an algebraic equation is generated after selecting an observation point $\vec{\xi}^l$ and performing collocation along the $N$ nodal points $\vec{x}^k$ leading to;

\begin{equation} \label{discrete BEM}
C_{ij}U^l_i=G_{ij}^{kl}t_i^k - H_{ij}^{kl}U_i^k.
\end{equation}

In \cref{discrete BEM} the terms $G_{ij}^{kl}$ and $H_{ij}^{kl}$ correspond to the integrals of the Green's functions over the $\vec{x}^k$-element and evaluated at the $\vec{\xi}^l$ collocation point.  Similarly, in the case of the indirect algorithm we generate an algebraic equation after selecting an observation point $\vec{x}^k$ and performing collocation along the $\vec{\xi}^l$   nodal points leading to;

\begin{equation} \label{discrete IBEM}
\begin{aligned}
U_i^k&=\tilde{G}_{ij}^{kl}\phi_j^l\\
t_i^k&=\tilde{H}_{ij}^{kl}\phi_j^l
\end{aligned}
\end{equation}

\begin{figure}[H]
\centering
\includegraphics[width=5.2 in]{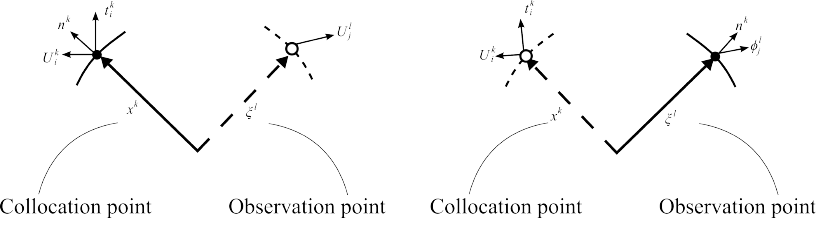}
\caption{Definition of the observation and collocation points in the
direct (left) and indirect (right) boundary element algorithms.}
\label{fig:collocation scheme}
\end{figure}

In \cref{discrete IBEM} the terms $\tilde{G}_{ij}^{kl}$ and $\tilde{H}_{ij}^{kl}$ have the same meaning as in the direct algorithm, however, it should be noticed that the observation and field points are now reversed.  Observe also that the displacement matrices $G_{ij}^{kl}$ and $\tilde{G}_{ij}^{kl}$ are the same in both methods because of the symmetries in the Green's function.  On the contrary, the traction matrices have both directional and evaluation arguments which are transposed.  The corresponding collocation scheme in each case is described in \cref{fig:collocation scheme}.

\begin{figure}[H]
\centering
\includegraphics[width=3.5 in]{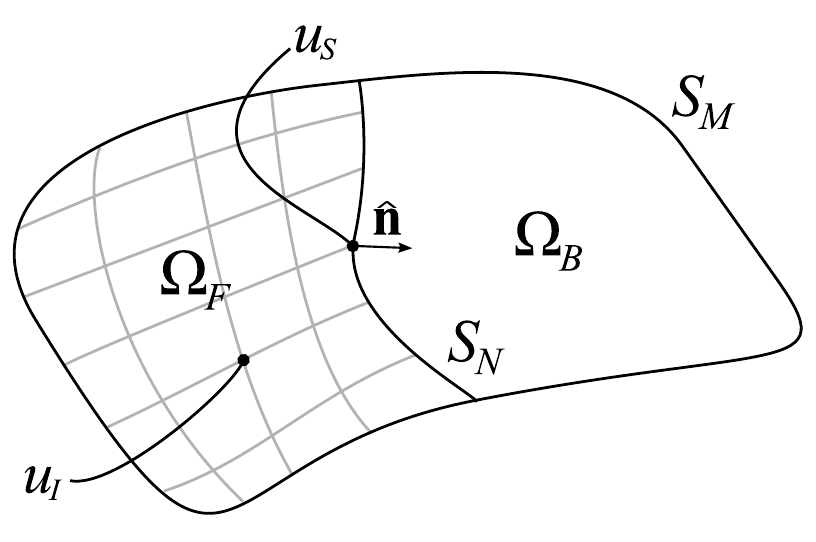}
\caption{Schematic description of the coupling between a BEM and a FEM discretization.}
\label{fig:coupling}
\end{figure}

The problem of coupling the BEM and FEM discretizations, reduces now to the simple exercise of writing the BEM equations--relating unknown nodal tractions and displacements to prescribed nodal tractions and displacements--into the equivalent form of a nodal forces-displacements relationship like in a standard displacements based finite element method formulation. Moreover, in the case of the discussed wave propagation problem, formulated in terms of the scattered field over the half-space and when the used Green's functions are those of a semi-infinite medium, the problem simplifies even more, since there are no remote forces originating at the far surface of the BEM domain, labeled $S_M$ in \cref{fig:coupling}, to be carried over to the FEM domain.

In the case of plane wave incidence, the stiffness relation to be coupled and directly implemented into an existing FEM code, reduces to the Half-Space-Super-Element (HSSE) stiffness matrix and its associated loads vector containing the incoming field, but expressed in terms of consistent nodal forces , see Bielak and Christiano (1984).  Referring again to \cref{fig:coupling}, we let the finite element part of the domain be $\Omega_F$. This domain is assumed to have been discretized into internal degrees of freedom $U_I$ and the contact surface into degrees of freedom $U_S$. In general, we will assume that the discretizations of the BEM and the FEM meshes have different number of nodes and therefore compatibility must be enforced through displacement and coupling matrices $R_u$ and $R_t$ \citep{tassoulas1988dynamic}, as specified in \cref{coupling Matrices};

\begin{equation} \label{coupling Matrices}
\begin{aligned}
U_N&=R_u U_S\\
F_N(\hat{n}^*)&=R_t t_S(\hat{n})
\end{aligned}
\end{equation}

and where the normal vectors satisfy $\hat{n}=-\hat{n}^*$. Now, we can write the discrete finite element equations corresponding to the scatterer like;

\begin{equation} \label{scatterer}
\begin{bmatrix}
  K_{II} & K_{IS} \\
  K_{SI} & K_{SS}
 \end{bmatrix}
 \begin{bmatrix}
   U_{I}\\
  U_{S}
 \end{bmatrix}
 =
\begin{bmatrix}
  F_I\\
  F_{S}(\hat{n})
 \end{bmatrix} 
\end{equation}

where $F_S(\hat{n})$ are the consistent nodal forces along the contact surface.  Similarly the equations for the half-space, after having converted the BEM system into a generalized force displacement relationship are written like;

\begin{equation} \label{HSSE}
F_N^S(\hat{n}^*)=K_{HS}U_S^S
\end{equation}

and where the forces along the contact surface have been denoted like $F_N^S(\hat{n}^*)$. Coupling of \cref{scatterer} and \cref{HSSE} using the conditions along the contact surface between displacements and forces involving the incoming field represented by the discrete terms $U^0_{S}$ and $F^0_{S}(\hat{n}^*)$ and given by;

\begin{equation}\label{contact conditions}
\begin{aligned}
U_{S}&= {U^S_{S}}+{U^0_{S}}\\
F_{S}(\hat{n})&+{F^S_{S}(\hat{n}^*)}+{F^0_{S}(\hat{n}^*)}=0
\end{aligned}
\end{equation}

\noindent
yields;

\begin{equation} \label{complete}
\begin{bmatrix}
  K_{II} & K_{IS} \\
  K_{SI} & K_{SS}+K_{HS}
 \end{bmatrix}
 \begin{bmatrix}
   U_{I}\\
  U_{S}
 \end{bmatrix}
 =
\begin{bmatrix}
  F_I\\
  -F_{S}^0(\hat{n}^*)+K_{HS}U_S^0
 \end{bmatrix} 
\end{equation}

In \cref{complete} the terms $U_S^0$ and $F_{S}^0(\hat{n}^*)$ represent the incoming field in terms of displacements and forces, evaluated at the contact surface and   $K_{HS}$ describes the half-space contribution to the global equilibrium equations. Denoting the Green tractions matrix for the direct boundary element and indirect boundary element method like $H^{BEM}$ and $H^{IBEM}$ respectively, yields after direct comparison of the resulting stiffness matrices the following connection between both algorithms

\begin{equation} \label{BEM AND IBEM}
\begin{aligned}
K_{HS}^{BEM}&=R_t H^{BEM} G^{-1} R_u\\
K_{HS}^{IBEM}&=R_t G^{-1} H^{IBEM}  R_u
\end{aligned}
\end{equation}

\subsection{Compression of the stiffness matrix}
The non-local nature of the BEM matrices $G$ and $H$ in \cref{BEM AND IBEM}, results in a finite element system of equations with a dense coefficient matrix. This result, associated to the non-local nature of the half-space superelement, is disadvantageous for its coupling in a finite element algorithm since the HSSE now controls the memory requirements and avoids the use of available and robust iterative solvers based on banded stiffness matrices. In this work we used two compression methods to impose a banded condition in the $HSSE$ stiffness matrix, so we can increase computer resources and take advantage of iterative solvers available in most commercial codes.  Although several works have been performed related to compression of BEM equations, we directly operate here on the resulting stiffness matrix.  To the best of our knowledge a study of how this manipulation affects accuracy has not been developed.  Two compression methods were used and defined as follows:

\begin{compactitem}
\item A threshold method, where all the terms smaller than a preselected percentage of the maximum absolute value present in the original matrix are made zero
\item A half-band-width method, where all the values located outside of a preselected distance from the diagonal are eliminated.
\end{compactitem}

\begin{figure}[H] 
\centering
\includegraphics[width=5.0 in]{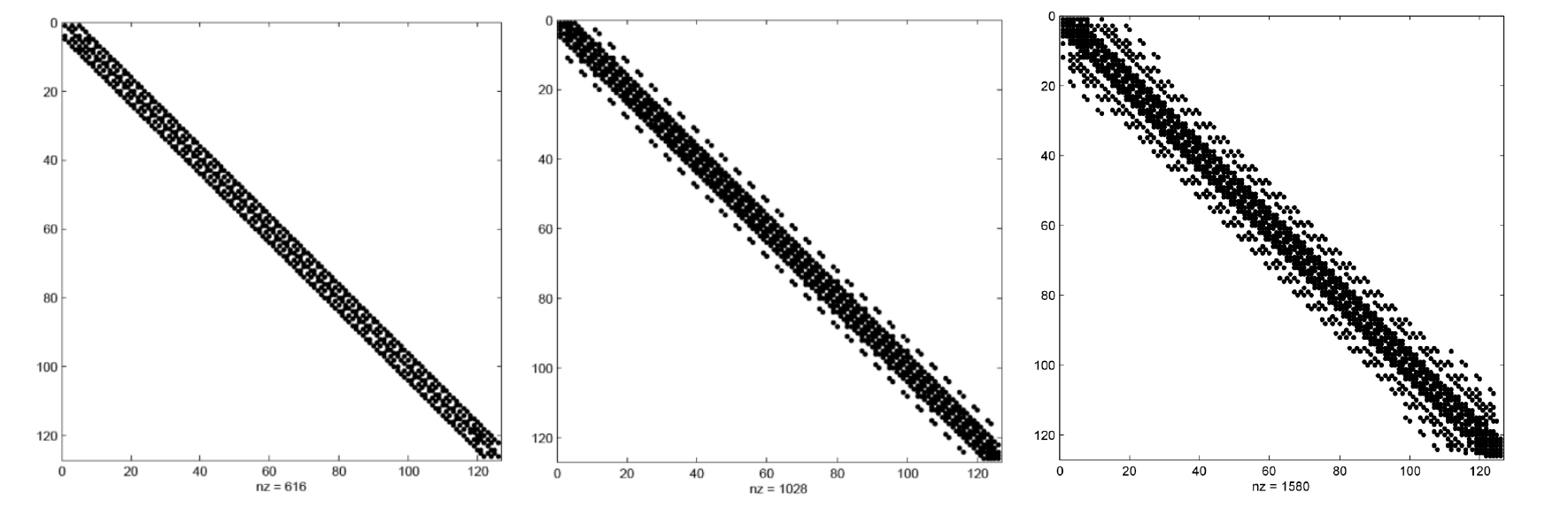}
\caption{Appearance of the stiffness matrices for a semicircular canyon after applying the threshold criteria with values (a) 0.01 (less accurate), (b) 0.001 and (c) 0.0001 (more accurate).}
\label{fig:threshold}
\end{figure}

Both methods are based on the tendency of the BEM matrices to be diagonally dominant due to the singularities in the Green's functions.  The threshold criteria however, is convenient as it can be applied with a sound physical basis, since the terms in the BEM matrices can be understood like influence coefficients indicating how strong is the effect of a given source in the response at a given nodal point; therefore the preselected threshold value is in fact a source intensity coefficient.  The trend of the matrices to be diagonally dominant can be observed in \cref{fig:threshold}, where we plotted the appearance of the reduced BEM stiffness matrices for the model of a semi-circular canyon after applying the threshold criteria. The BEM stiffness matrix can be computed and compressed inside a user element subroutine, coupled to an existing finite element code. It suffices  to specify the nodal points conforming the contact surface $S_M$ and the subroutine must return the elemental contribution (see \cref{complete} and \cref{BEM AND IBEM}). If the problem is directly solved in the time domain, it suffices to implement a user element subroutine UEL where the element contribution to the global system is either one of the $K$ 's specified in \cref{BEM AND IBEM} and the contribution to the loads vector is the one appearing in \cref{complete}. On the other hand, if the analysis is performed in the frequency domain, where the $K$'s are complex valued, one needs to assume that the model has real and imaginary degrees of freedom at each node, thus doubling the number of degrees of freedom with respect to the ones used in a direct time domain analysis.  With this assumption at hand, the implementation of the finite element proceeds exactly like in the time domain. For instance, such a generalized framework has been proposed by \cite{mosquera2013implementation}. 

\section{Results}
\subsection{Compressed Stiffness Matrix Formulation}
The well known benchmark problems of a semi-circular and a rectangular canyon were selected in order to test and evaluate the loss of accuracy of the stiffness matrix approximation. These problems constitute a worst case scenario for the approximation if the resulting matrices are interpreted like absorbing boundaries. The response of both canyons to $P$ and $SV$ waves incident at $0^\circ$ and $30^\circ$ with respect to the vertical was calculated. For the excitation we applied an incoming field corresponding to a Ricker pulse. As a threshold value we applied a variation from 0.0 to 0.01, with 0.0 corresponding to the original complete matrix and with a larger value representing a less accurate approximation. In the case of the band-width criteria the size of the relative half-band width was varied from 0.03 to 1.0 with the complete matrix corresponding to a value of 1.0 and a null value to a matrix with only the diagonal terms. The values of the relative half-band width for matrices approximated with the threshold criteria are shown in \cref{tab:semiancho}.

\begin{table}[H]
\centering
\begin{tabular}{ccccc}
\hline 
\multirow{2}{*}{\textbf{Threshold}} & \multicolumn{2}{c}{\textbf{Relative half-bandwidth}}  & \multicolumn{2}{c} {\textbf{Relative storage}}\\ 
 & \textbf{Semicircular} & \textbf{Rectangular}
 & \textbf{Semicircular} & \textbf{Rectangular} \\ 
\hline
 $0$ & $1$ & $1$  & $100\%$ & $100\%$  \\ 
 $10^{-5}$ & $0.26$ & $0.25$ & $45.2\%$ & $43.8\%$ \\ 
 $10^{-4}$ & $0.13$ & $0.09$ & $24.3\%$ & $17.2\%$ \\ 
 $10^{-3}$ & $0.06$ & $0.05$ & $11.6\%$ & $9.8\%$ \\ 
 $10^{-2}$ & $0.03$ & $0.02$ & $5.9\%$ & $4.0\%$ \\ 
\hline 
\end{tabular} 
\caption{Relative half-band-width for the dynamic matrices compressed by the threshold criteria. Relative storage values are also presented --storing just the terms in the band.}
\label{tab:semiancho}
\end{table}

After imposing the banded condition we are interested in determining the storage requirements for the new resultant matrix.  The original needs corresponding to the full matrix requires $n^2$ memory words.  For the new approximated matrix and storing just the terms in the band, the relative needed storage requirement, $Rst$, can be approximated after neglecting the words required to store the diagonal itself like;

\begin{equation} 
Rst(Rhbw)=2Rhbw-Rhbw^2
\label{MEMORY}
\end{equation}

and where $Rhbw$ is the relative half-band width.  A storage requirement corresponding to $50\%$ of the full matrix is equivalent to a relative half-band width of 0.3, while 0.13 would be equivalent to $25\%$ of the original storage requirement.

\begin{figure}[H]
\centering
\includegraphics[width=5.5 in]{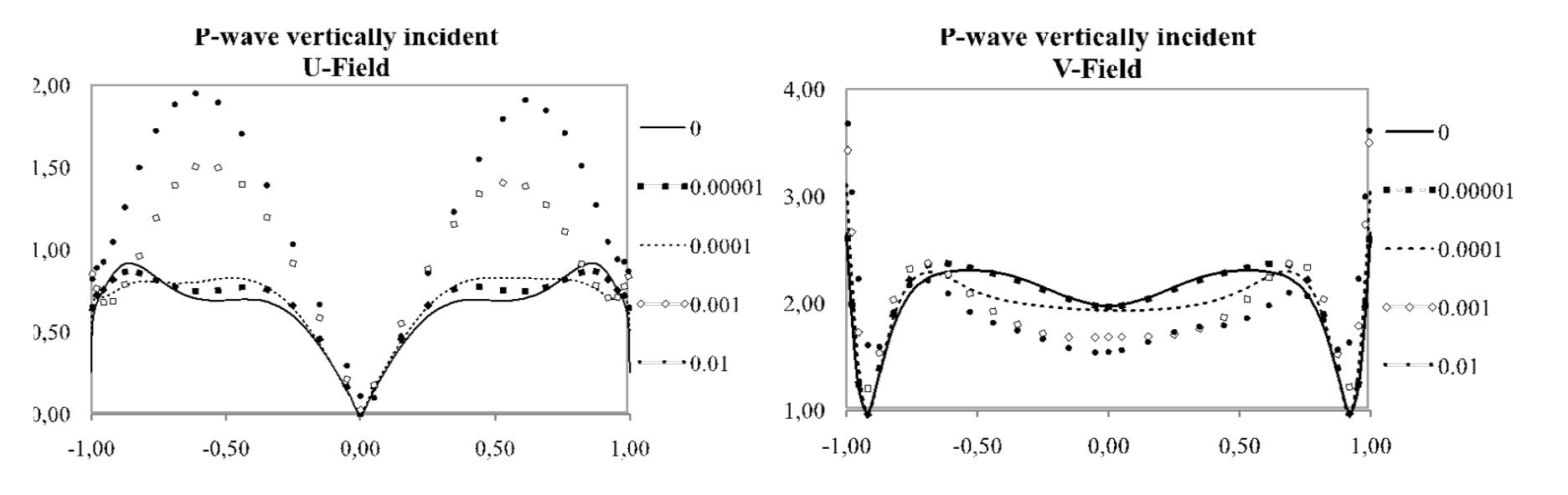}
\caption{Transfer function for a semicircular canyon for a vertically incident P wave at a dimensionless frequency of 1.}
\label{fig:TFSC1}
\end{figure}

\begin{figure}[H]
\centering
\subfloat[Complete matrix] {\includegraphics[width=2.2 in]{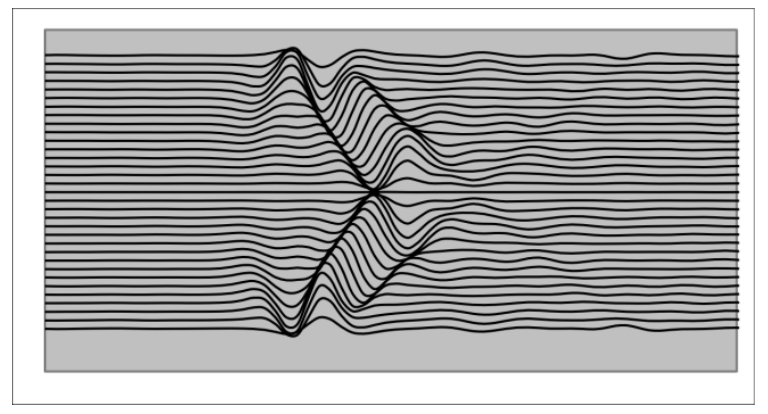}}
\subfloat[Threshold value=0.00001] {\includegraphics[width=2.2 in]{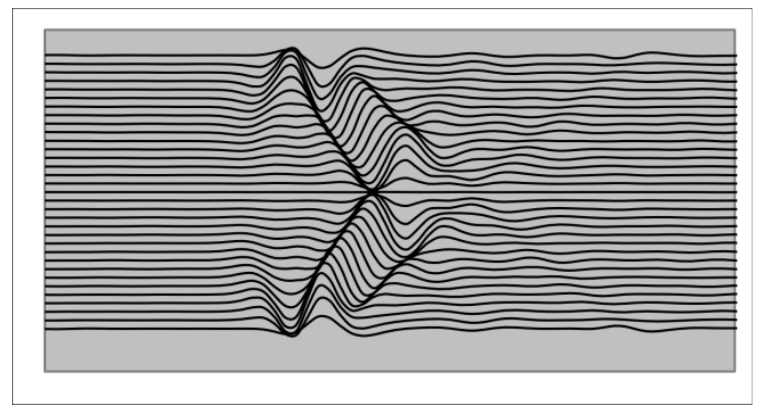}}\\
\subfloat [Threshold value=0.0001] {\includegraphics[width=2.2 in]{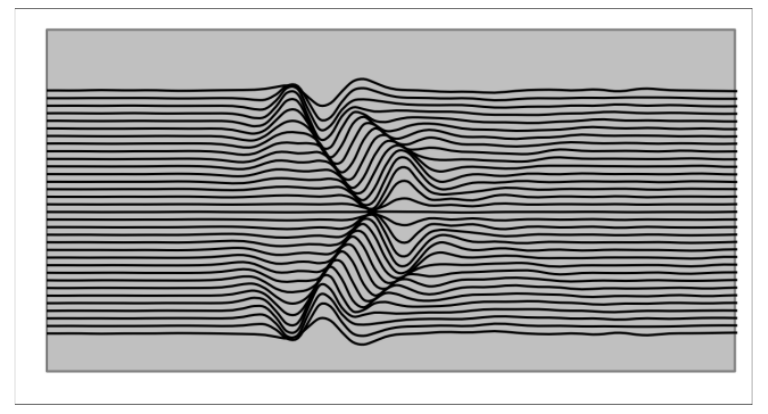}}
\subfloat [Threshold value=0.01] {\includegraphics[width=2.2 in]{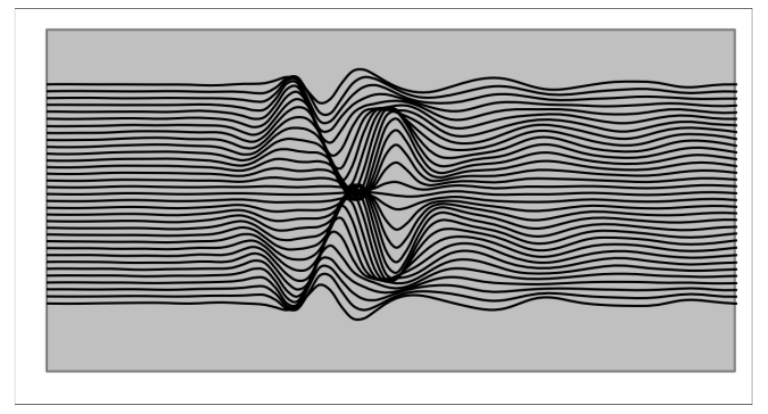}}
\caption{Synthetic seismograms for horizontal displacement for a semicircular canyon for a vertically incident P wave for different threshold values.}
\label{fig:SYNC1}
\end{figure}

\begin{figure}[H]
\centering
\includegraphics[width=5.5 in]{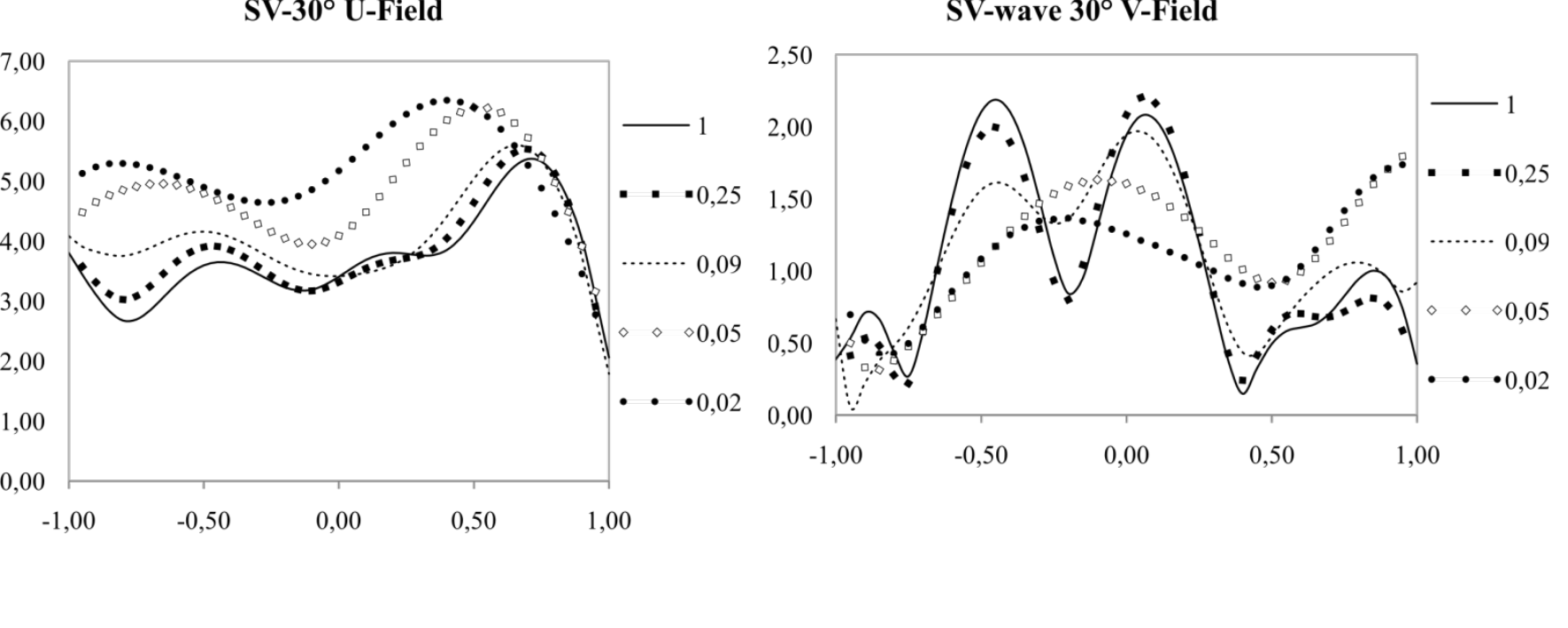}
\caption{Transfer function for a rectangular canyon for a $30^\circ$ incident SV wave at a dimensionless frequency of 1.}
\label{fig:TFRE1}
\end{figure}

\begin{figure}[h]
\centering
\subfloat[Complete matrix] {\includegraphics[width=2.2 in]{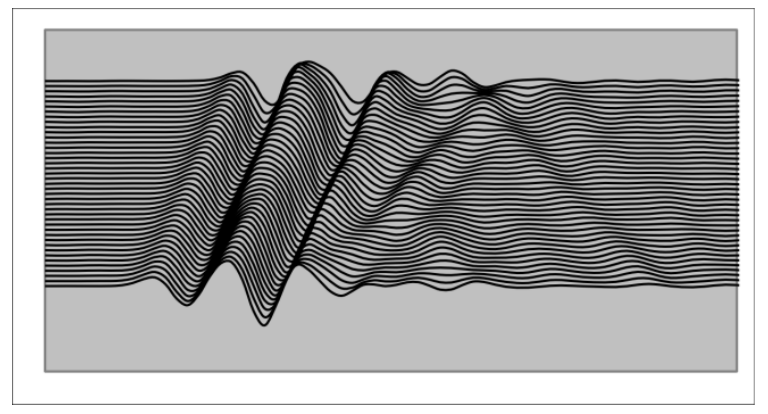}}
\subfloat[Threshold value=0.00001] {\includegraphics[width=2.2 in]{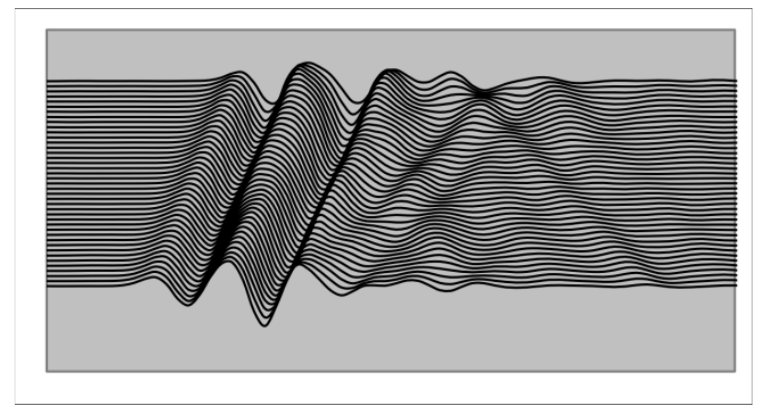}}\\
\subfloat [Threshold value=0.0001] {\includegraphics[width=2.2 in]{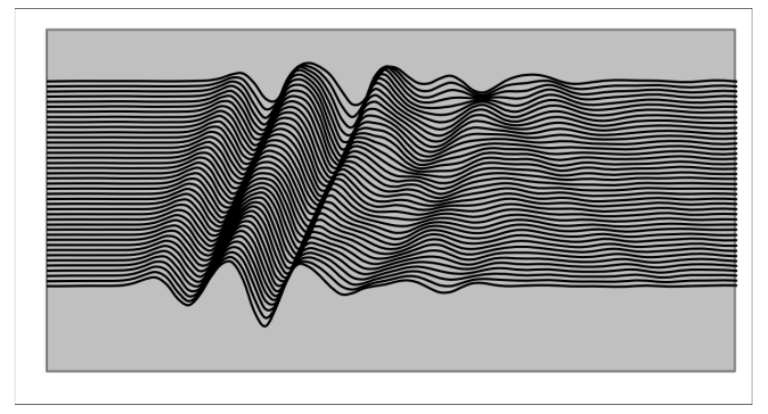}}
\subfloat [Threshold value=0.01] {\includegraphics[width=2.2 in]{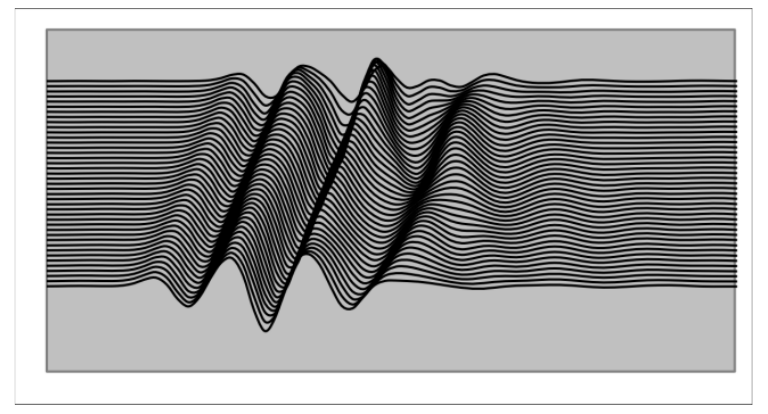}}
\caption{Synthetic seismograms for horizontal displacement for a rectangular canyon for a vertically incident SV wave for different threshold values.}
\label{fig:SYNC2}
\end{figure}

\begin{figure}[H]
\centering
\includegraphics[width=5.5 in]{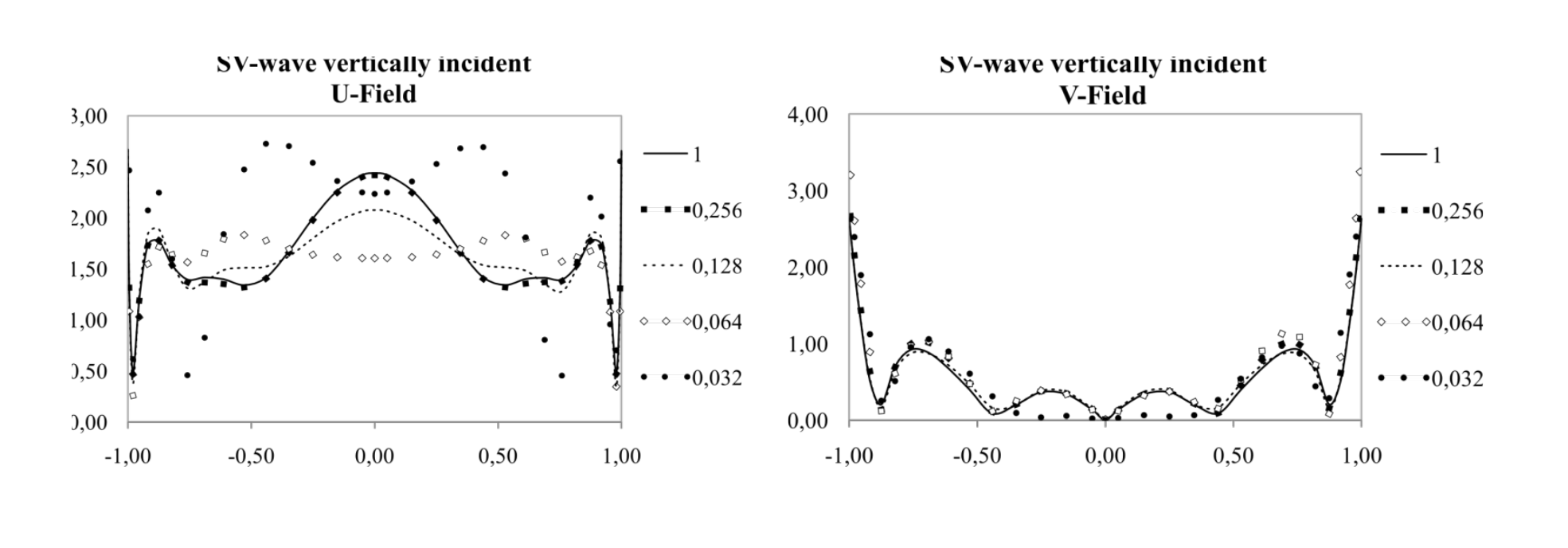}
\caption{Transfer function for a semicircular canyon for a vertically incident SV wave at a dimensionless frequency of 1.}
\label{fig:TFSC2}
\end{figure}

\begin{figure}[H]
\centering
\subfloat[Complete matrix] {\includegraphics[width=2.2 in]{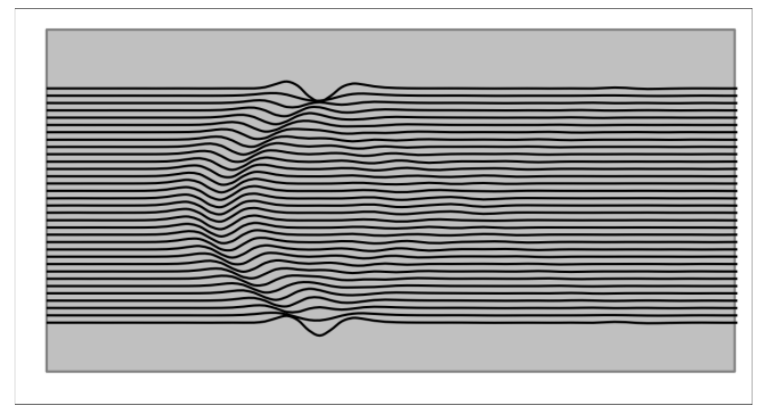}}
\subfloat[Relative half-bandwidth=0.26] {\includegraphics[width=2.2 in]{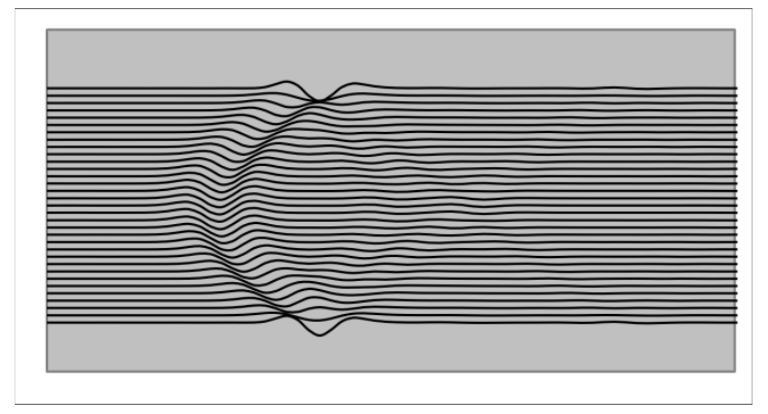}}\\
\subfloat [Relative half-bandwidth=0.13] {\includegraphics[width=2.2 in]{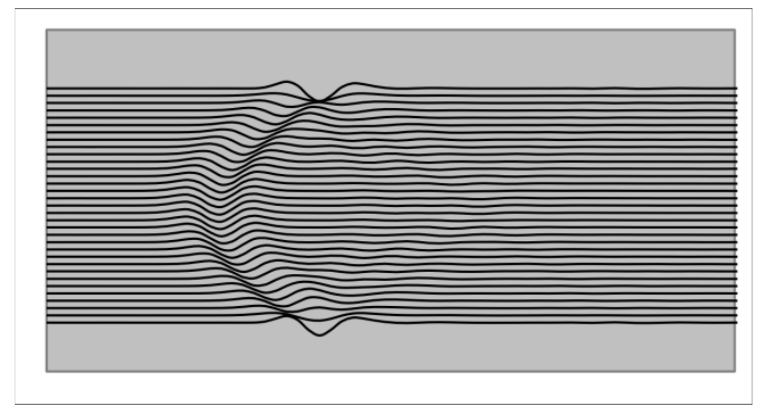}}
\subfloat [Relative half-bandwidth=0.03] {\includegraphics[width=2.2 in]{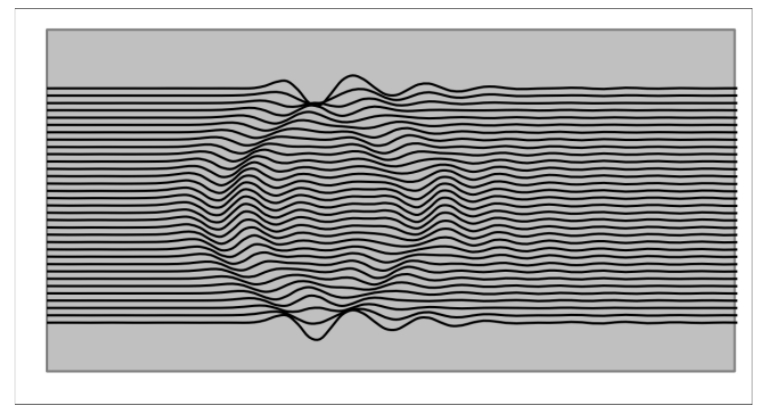}}
\caption{Synthetic seismograms for horizontal displacement for a semicircular canyon for vertically incident SV wave different half-bandwidth values.}
\label{fig:SYNC3}
\end{figure}

\begin{figure}[H]
\centering
\includegraphics[width=5.5 in]{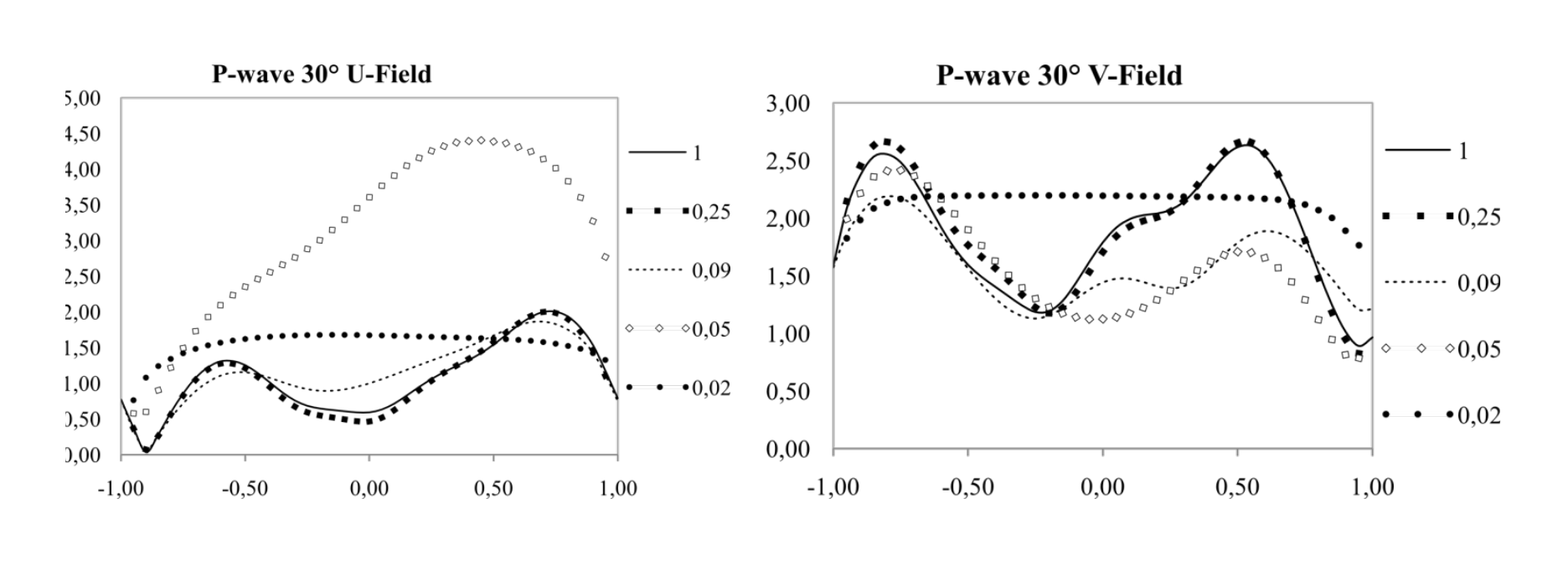}
\caption{Transfer function for a rectangular canyon for a $30^\circ$ incident P wave at a dimensionless frequency of 1.}
\label{fig:TFRE2}
\end{figure}

\begin{figure}[H]
\centering
\subfloat[Complete matrix] {\includegraphics[width=2.2 in]{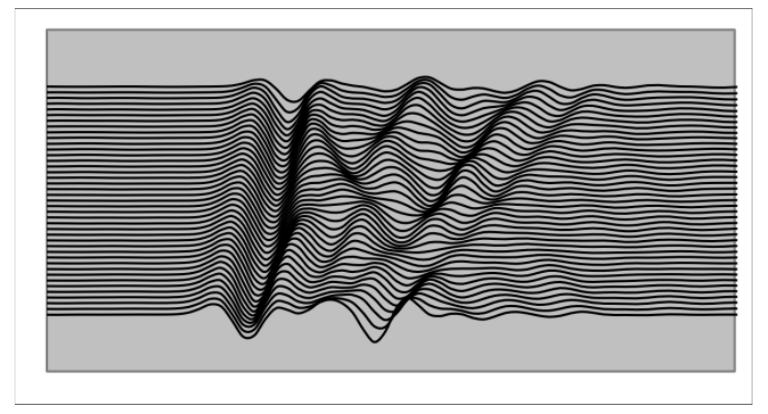}}
\subfloat[Relative half-bandwidth=0.25] {\includegraphics[width=2.2 in]{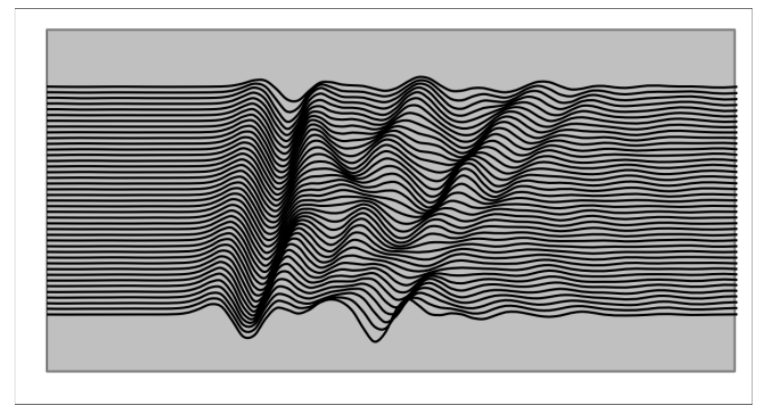}}\\
\subfloat [Relative half-bandwidth=0.09] {\includegraphics[width=2.2 in]{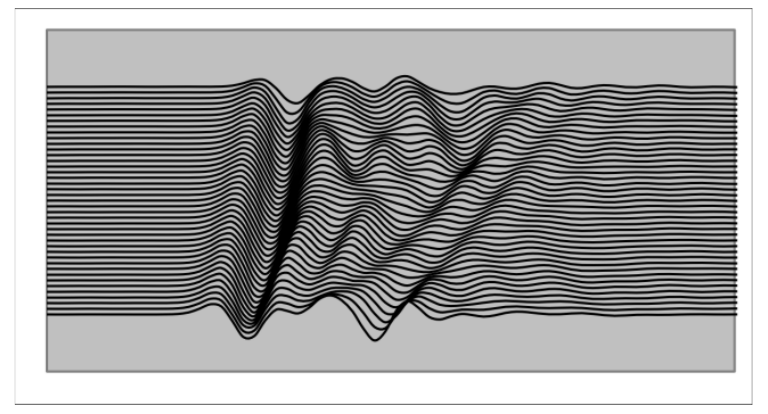}}
\subfloat [Relative half-bandwidth=0.02] {\includegraphics[width=2.2 in]{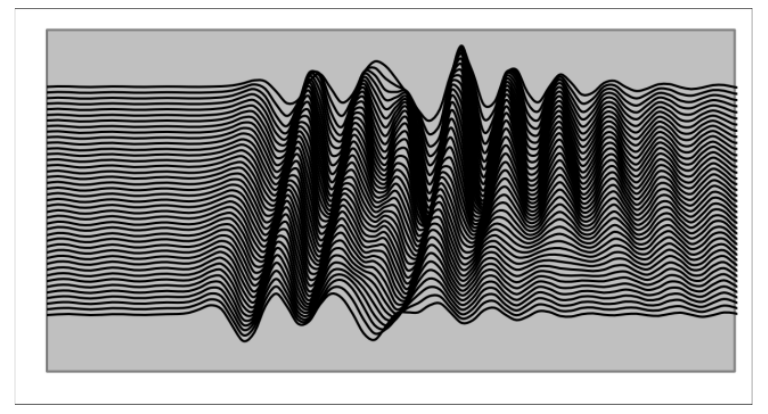}}
\caption{Synthetic seismograms for horizontal displacement for a semicircular canyon for a $30^\circ$ incident P wave for different half-bandwidth values.}
\label{fig:SYNC4}
\end{figure}

\begin{figure}[H]
\centering
\subfloat[Semicircular canyon] {\includegraphics[width=3 in]{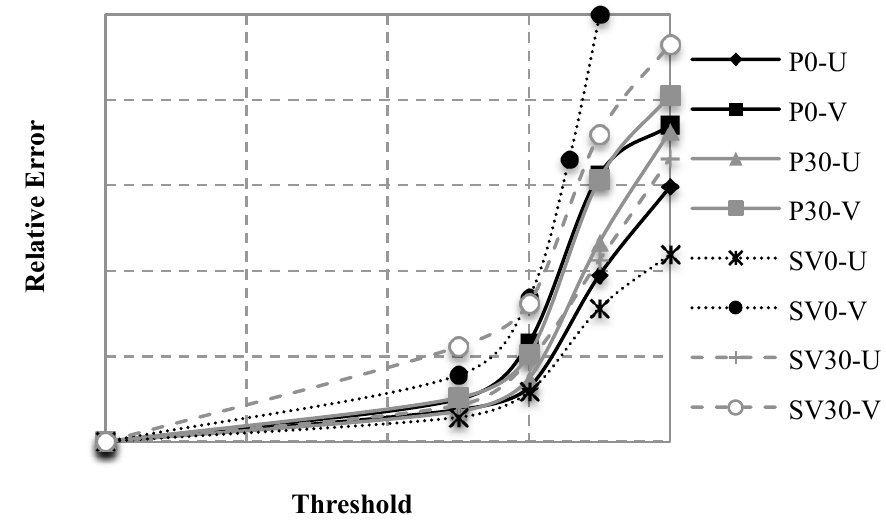}}
\subfloat[Rectangular canyon] {\includegraphics[width=3 in]{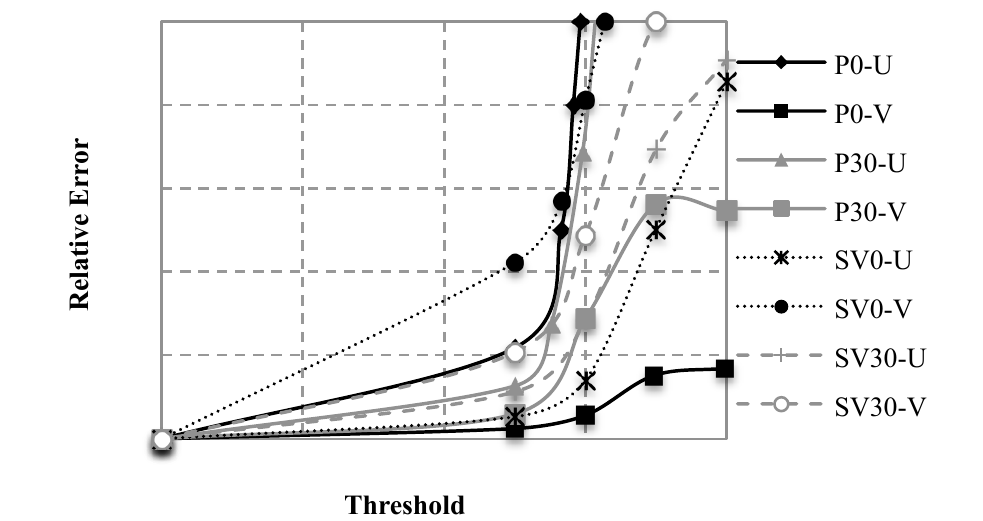}}\\
\subfloat [Semicircular canyon] {\includegraphics[width=3 in]{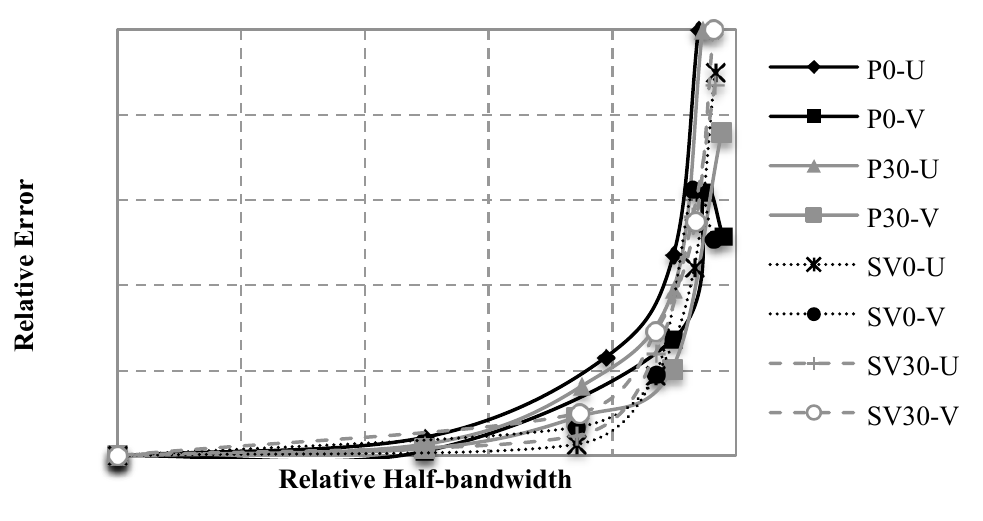}}
\subfloat [Rectangular canyon] {\includegraphics[width=3 in]{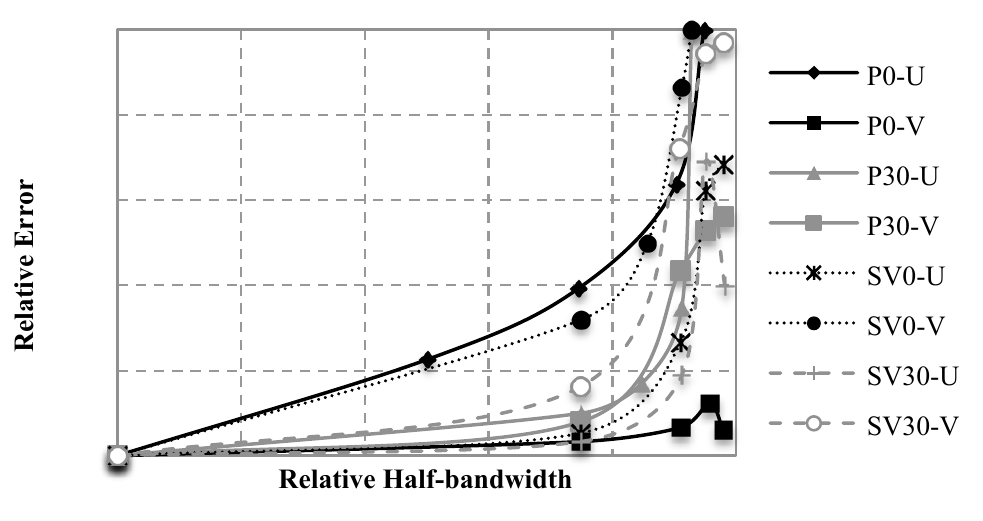}}
\caption{Comparison for different matrix compression criteria: a), b) for Threshold criterion and c), d) for Half-bandwidth criterion. In all the cases the geometry and the incidence type of wave SV or P and angle is varied for well known cases. The error is computed for the transfer function at a dimensionless frequency of 1.}
\label{fig:COMP CRIT}
\end{figure}

\Crefrange{fig:TFSC1}{fig:SYNC4} show the transfer functions and synthetic displacement time-histories (synthetic  seismograms) for a dimensionless frequency $\eta=\omega L/\pi \beta=1.0$ where $\omega$ is the circular frequency, $\beta$ the shear wave velocity and $L$ is the characteristic canyon dimension for different cases. On the other hand we show in \cref{fig:COMP CRIT} the relative errors with respect to the square-norm for different levels of approximation and for all considered incident waves.  In general, the best results were obtained for the vertical displacements under $P-$wave incidence and for the horizontal displacements in the case of $SV-$wave incidence.  At the same time, as was originally expected better results are obtained for the semi-circular canyon which contains less diffraction sources as compared to the rectangular canyon. This behaviour is observed for both the threshold and half-band-width criteria.

\section{Conclusions and Furtherwork}
\label{sec:conclu}
Two approximate numerical solution methods for the elastic wave scattering problem, were discussed and evaluated in order to facilitate the study of dispersion problems at the practising engineering level in available personal computers and within existing finite element architectures.  First, the loss of accuracy associated with two different ways of converting fully dense stiffness matrices into banded stiffness matrix was evaluated.  The original stiffness matrix appeared after the discretization of a half-space sub-domain with two different boundary element algorithms, e.g., a direct boundary element method and an indirect boundary element method.  The proposed techniques are oriented to the solution of the problem at the engineering practising level since one arrives to a stiffness matrix that can be coupled into existing commercial finite element codes, e.g., ABAQUS and FEAP.  The compression procedure results in savings in memory requirements.  Although similar procedures have been previously used in order to compress full BEM matrices, to the best of our knowledge there are no published works for the compression of coupled BEM/FEM algorithms.  A good approximation is obtained for vertical displacements under P-wave incidence and horizontal displacements under SV-wave incidence.  A larger error is obtained for the rectangular canyon because of the presence of singularity points in the tractions field along the corners.  Errors as large as $50\%$ for the considered range of approximations were found.  For an error of the order of $10\%$ a relative half-band-width of 0.13 or equivalently, a relative storage requirement of 0.25 can be used.  This is a memory saving of $75\%$.

\subsection*{Aknowledgements}
This project was conducted with financial support from ``Departamento Administrativo de Ciencia, Tecnolog\'ia e Innovaci\'on, COLCIENCIAS' and from Universidad EAFIT.

\nocite{MSc_thesis-Guarin2012}
\bibliographystyle{gji}
\bibliography{references}
\end{document}